\RequirePackage{lineno} 
\documentclass[aps,prb,reprint]{revtex4-1}

\usepackage{bm}%
\usepackage{dcolumn}
\usepackage{graphicx}
\usepackage{float}
\usepackage{soul}
\usepackage{amsmath}
\usepackage{url}

\begin{document}

\title{Calculation of currents around a triangular indentation by the hodograph method}

\author{J. I. Avila$^1$,  B. Venderheyden$^2$, A. V. Silhanek$^3$, S. Melinte$^1$}

\address{$^1$ ICTEAM, Universit\'e catholique de Louvain, B-1348, Louvain-La-Neuve, Belgium}
\address{$^2$ SUPRATECS and Department of Electrical Engineering and Computer Science, Universit\'e de Li\`ege, B-4000 Sart Tilman, Belgium}
\address{$^3$ Experimental Physics of Nanostructured Materials, Q-MAT, CESAM, Universit\'e de Li\`ege, B-4000 Sart Tilman, Belgium}

\date{\today}

\begin{abstract}
Border indentations in non-linear conductors, such as superconducting thin films in the creep regime, alter the distribution of currents and magnetic fields near and far from the indentation. One of such disturbances are the discontinuity lines, or \textit{d}-lines, a parabolic-like line originating from the indentation where the current density direction changes abruptly. Hodograph series results are obtained for the currents around a triangular indentation and its corresponding $d$-lines in a conducting stripe of finite width and in an infinite half plane, considering two cases: uniform creep exponent and mixed infinite and ohmic exponents. The mixed creep exponent case presents currents distributions resembling the purely ohmic case, with significant current disturbances only near the indentation. For uniform creep exponent, results similar to a planar indentation are obtained, with far ranged currents features  and  parabolic-like $d$-lines with shapes depending on the creep exponent. In particular, the same $d$-line asymptotic behaviour is obtained for the triangle indentation as that of the planar defect in the critical state, a result obtained here just on continuity considerations of the hodograph expansions. This equivalence is due to identical contributions to the Fourier series of the current stream-function in the hodograph space, obtained from an images method expansion.
\end{abstract}

\pacs{...}

\maketitle

\section{Introduction}
Defects have significant effects on the electrical properties of type II superconductor. While nanoscopic defects serve as pinning centres that arrest the motion of flux vortices and avoid dissipation \cite{Brandt0, Koshelev, Blatter}, larger defects create zones of increased resistivity that can serve as nucleation spots for flux avalanches \cite{Jerem1, Mints}. Border defects such as cracks or voids are accompanied by long range perturbations in the current density and magnetic field similar to the Bean model \cite{Bean1,Bean2,Bean3} discontinuity lines \cite{Schuster1,Schuster2, Eisenmenger}, where the current density $J$ turns sharply. The exact shape of the lines will be directly related to the critical current and constitutive relationship under external applied fields, such as an external electric field, or an external magnetic field ramp creating induction currents. The resistivity in the flux creep regime \cite{Vinokur1, Mints} can be modelled as: 
\begin{equation}
\rho(J,B,T)=\rho_0 (J/J_\mathrm{c}(B,T))^{n-1},
\label{ExpCreep}
\end{equation}
where $\rho_0$ is the resistivity at $J=J_\mathrm{c}$,  $B$ the magnetic field, $T$ the temperature and $n$ is the creep exponent. The steep variations of the resistivity in this regime are also espected to generate thermo-magnetic instabilities \cite{Mints, Rak, Deni}. By taking no magnetic field or temperature dependence of $J_c$, expression  (\ref{ExpCreep}) has been used by means of the hodograph method \cite{Gur1,Gur2,Gur3} to obtain analytical expressions for the electric currents around a planar defect, for finite and infinite creep exponent $n$. A more basic model consists of considering only the case $n\rightarrow\infty$ and to disregard non-local effects and magnetic field or temperature depence of the critical current. This leads to the Bean model \cite{Bean1,Bean2,Bean3, Camp1}, in which the magnetization current density of a sample exposed to a ramping magnetic field is modelled as piece-wise uniform regions, on whose boundaries the current can change of direction discontinuously, tracing the $d$-lines. From this, parabolic \textit{d}-lines of the form $x = y^2/2h+h/2$ are obtained for an indentation of height $h$ on a border located at $x=0$ \cite{Jerem1}. Here, the calculations of currents around a planar defect by the hodograph method \cite{Gur2,Gur3} are extended to a triangular defect. In that formalism the calculated currents actually do not change direction in a discontinuous way, however, the $d$-line concept is extended to a current domain wall that traces a line similar to that of the Bean's model, so the $d$-line term is also used here. The current stream function is obtained for different creep exponents and geometries and finally compared to the Bean model.

\section{Hodograph method}
A general stationary solution method was developed in refs \cite{Gur1,Gur2,Gur3} for currents in films with a non-linear relationship between the electric field $E$ and the current $J$ of the type $E=E_0 (J/J_0)^n$, where $E_0$ is a characteristic electric field magnitude and $J_0$ the corresponding current density and $n\geq 1$. In this method the Maxwell equations for the current stream function $\psi$ are solved in the space of the current density magnitude $J=\vert\nabla \times \psi\vert$ and the angle $\theta$ subtended between $\bm{J}$ and the $x$ axis, i.e. under the transformation $\psi (x,y)\rightarrow \psi (J,\theta)$. It can be equally applied for the electric potential with $\phi (x,y) \rightarrow \phi (E,\theta)$, where $E=-\nabla \phi$. In the space $(J,\theta)$, namely the hodograph space, the electromagnetic equations become linear and separable when using the power law resistivity, and can be solved by eigenfunction expansions, with general solution:
\begin{small}
\begin{multline}
\frac{\psi(J,\theta)}{I}= A_0+(B_0+D_0 \theta)\left(\frac{J}{J_0}\right)^{(1-n)}+C_0 \theta \\ + \sum_{m=1}^{\infty} C_{m}\left( \frac{J}{J_0}\right)^{\tau_{m}^+}\sin (m\theta+\phi_m) \\ + \sum_{m=1}^{\infty} D_{m}\left( \frac{J}{J_0}\right)^{\tau_{m}^-}\sin (m\theta+\omega_m), \label{psi}
\end{multline}
\end{small}
with
\begin{equation}
\tau_{m}^\pm=\frac{1}{2}(1-n\pm \sqrt{(n-1)^2+4nm^2}).
\end{equation}
The boundary conditions for $\psi$ in the hodograph space are simple only in the case of samples with straight borders, in which $\theta$ is constant. Once the solution is found in the hodograph plane, it can be transformed back to the $x$, $y$ space, i.e. $(J,\theta) \rightarrow (x,y)$. By using the complex variable $z=x+iy$, this transformation is given by:
\begin{small}
\begin{multline}
\frac{z}{l}=F_0+C_0 e^{i\theta}\frac{J_0}{J}+\\+\frac{e^{i\theta}}{n}\left(\frac{J}{J_0}\right)^{-n}\{i B_0(n-1)+D_0 (1+i(n-1)\theta    \}+\\
-e^{i\theta}\sum_{m=1}^{\infty} \frac{D_{m}}{\tau_{m}^- -1}\left( \frac{J}{J_0}\right)^{\tau_{m}^- -1}\left[m\cos (m\theta+\omega_m) \right.\\ \left. -i {\tau_{m}^-} \sin(m\theta+\omega_m)\right]+\\
-\ e^{i\theta}\sum_{m=1}^{\infty} \frac{C_{m}}{\tau_{m}^+ -1}\left( \frac{J}{J_0}\right)^{\tau_{m}^+ -1}\left[m\cos (m\theta+\phi_m) \right.\\ \left. -i {\tau_{m}^+} \sin(m\theta+\phi_m)\right],\label{z}
\end{multline}
\end{small}
Additional details are given in the appendix. The technique has been applied to geometries like corners, bridges, current leads and planar defects. In ref. \cite{Gur2}  a planar defect of width $2a$ perpendicular to the current in an infinite superconductor was considered. In that work, the \textit{d}-line was identified as the core of a current domain wall starting near the defect. In the limit $n\rightarrow \infty$ it was found there that, asymptotically, the \textit{d}-line  traces a curve with $x\approx y^2 /1.94a$, similar but not identical to the \textit{d}-line in the Bean result that goes as $x= y^2 /2a$.

\begin{figure}
\centering \includegraphics[]{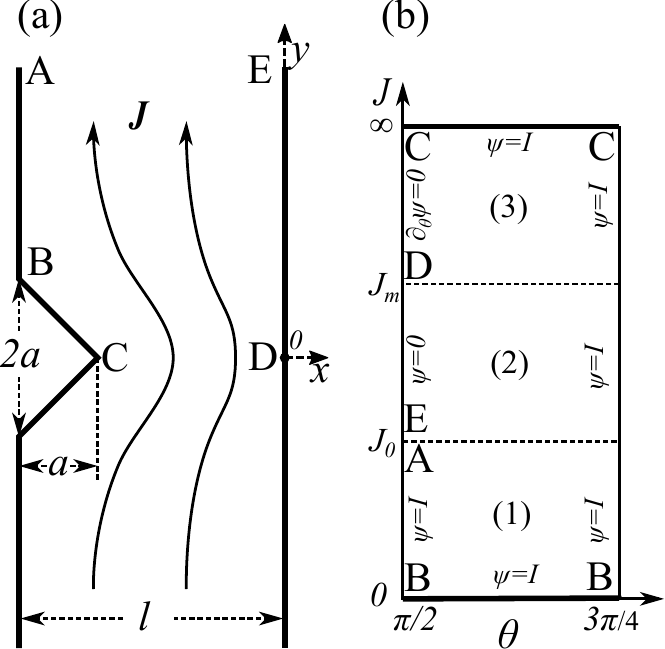}
\caption{(a) Geometry of superconducting strip with triangular constriction, indicating with capitals the points with direct equivalent in hodograph space. In particular the current density is zero in the corner B and reaches $J_m$ at $(0,0)$ and $J_0$ far from the constriction. The angle subtended by segment CB is $3\pi /4$. The stream function $\psi$ is chosen to be 0 on the right and $I$ on the left boundaries of the strip. Only the upper part $y\geq 0$ of the sample is solved, the remaining is obtained by symmetry. (b) Hodograph plane representation of solution for $\psi$ along with the boundary conditions following example of planar constriction in ref. \cite{Gur3}.}
\label{FigGur1}
\end{figure}
\section{Stripe of finite width}
Equations (\ref{psi}) and (\ref{z}) are solved for a strip carrying a total current $I$, of width $l$  and with a \textit{triangular} constriction of base $2a$ and height $a$, as depicted in Fig.\ref{FigGur1}(a). The solution for the stream function $\psi(J,\theta)$ is separated in 3 hodograph regions, as shown in Fig.\ref{FigGur1}(b), namely $\psi_1$ in region (1) for $0\leq J\leq J_0$, with $J_0$ the magnitude of $J$ far from the defect,  $\psi_2$ in region (2) for $J_0\leq  J\leq J_m$, where $J_m$ is the magnitude of the current density in the border directly opposite to the defect, and $\psi_3$ in region (3) for $J_m\leq J$. The solutions fulfilling the boundary conditions described in Fig.\ref{FigGur1}(b) are:
\begin{small}
\begin{equation}
\frac{\psi_1}{I}=1+\sum_{m=1}^{\infty} C_{4m}^{(1)}\left( \frac{J}{J_0}\right)^{\tau_{4m}^+}\sin (4m\theta),
\label{psi1}
\end{equation}
\end{small}

\begin{small}
\begin{multline}
\frac{\psi_2}{I}=-2+\frac{4\theta}{\pi}+\sum_{m=1}^{\infty} \left[ D_{4m}^{(2)}\left( \frac{J}{J_0}\right)^{\tau_{4m}^-} +\right.\\ \left. +C_{4m}^{(2)}\left( \frac{J}{J_m}\right)^{\tau_{4m}^+} \right] \sin (4m\theta),
\label{psi2}
\end{multline}
\end{small}

\begin{small}
\begin{equation}
\frac{\psi_3}{I}=1+\sum_{m=1}^{\infty}  D_{4m-2}^{(3)}\left( \frac{J}{J_m}\right)^{\tau_{4m-2}^-}  \sin ((4m-2)\theta+\frac{3\pi}{2}),
\label{psi3}
\end{equation}
\end{small}
The $C_{4m}^{(1)}$, $ D_{4m}^{(2)}$, $C_{4m}^{(2)}$ and $ D_{4m-2}^{(3)}$ coefficients are obtained by enforcing continuity of $\psi$ and $\partial \psi /\partial J$ at $J=J_0$ and $J=J_m$ along with Fourier analysis for $\theta \in[\pi/2,3\pi/4]$. 
Applying continuity at $J_0$:
\begin{small}
\begin{equation}
\frac{8}{\pi}\int_{\pi/2}^{3\pi/4}\psi_1(J_0)\sin(4m\theta)d\theta=\frac{8}{\pi}\int_{\pi/2}^{3\pi/4}\psi_2(J_0)\sin(4m\theta)d\theta
\end{equation}
\end{small}
It gets:
\begin{equation}
C_{4m}^{(1)}=-\frac{2}{m\pi}+D_{4m}^{(2)}+r^{\tau_{4m}^+}C_{4m}^{(2)}
\label{contJ0}
\end{equation}
Where $r=J_0/J_m$. From the derivative continuity at $J_0$, 
it is obtained:
\begin{equation}
C_{4m}^{(1)}=\frac{\tau_{4m}^-}{\tau_{4m}^+} D_{4m}^{(2)}+r^{\tau_{4m}^+}C_{4m}^{(2)}
\label{contdJ0}
\end{equation}
From continuity at $J_m$,
by using:
\begin{footnotesize}
\begin{equation}
\frac{8}{\pi}\int_{\pi/2}^{3\pi/4}\sin((4k-2)\theta+\frac{3\pi}{2})\sin(4m\theta)d\theta=\frac{8}{\pi}\frac{4m}{(4m)^2-(4k-2)^2}
\label{ecMixSines}
\end{equation}
\end{footnotesize}
It gets:
\begin{equation}
C_{4m}^{(2)}=\frac{2}{m\pi}-D_{4m}^{(2)}r^{-\tau_{4m}^-}+\frac{8}{\pi}\sum_{k=1}^{\infty}\frac{4mD_{4k-2}^{(3)}}{(4m)^2-(4k-2)^2}
\label{contJm}
\end{equation}
And from continuity of the derivative at $J_m$:
\begin{equation}
\tau_{4m}^- D_{4m}^{(2)}r^{-\tau_{4m}^-}+\tau_{4m}^+ C_{4m}^{(2)}=\frac{8}{\pi}\sum_{k=1}^{\infty}\frac{4mD_{4k-2}^{(3)} \tau_{4k-2}^-  }{(4m)^2-(4k-2)^2}
\label{contdJm}
\end{equation}
Combining (\ref{contJ0}) and (\ref{contdJ0}):

\begin{equation}
D_{4m}^{(2)}=\frac{2}{m\pi}\frac{\tau_{4m}^+}{\tau_{4m}^+ - \tau_{4m}^-}
\label{eqD2}
\end{equation}
And combining (\ref{contJm}), (\ref{contdJm}) and (\ref{eqD2}) the linear system of equations for $ D_{4m-2}^{(3)}$  is obtained:
\begin{equation}
\frac{\tau_{4m}^+}{(4m)^2}(r^{-\tau_{4m}^-}-1)=\sum_{k=1}^{\infty}\frac{D_{4k-2}^{(3)}(\tau_{4m}^+-\tau_{4k-2}^-)}{(4m)^2-(4k-2)^2}
\label{eqD3s}
\end{equation}

\begin{figure*}
\centering
\begin{minipage}{.5\textwidth}
  \centering
  \includegraphics[width=.74\linewidth]{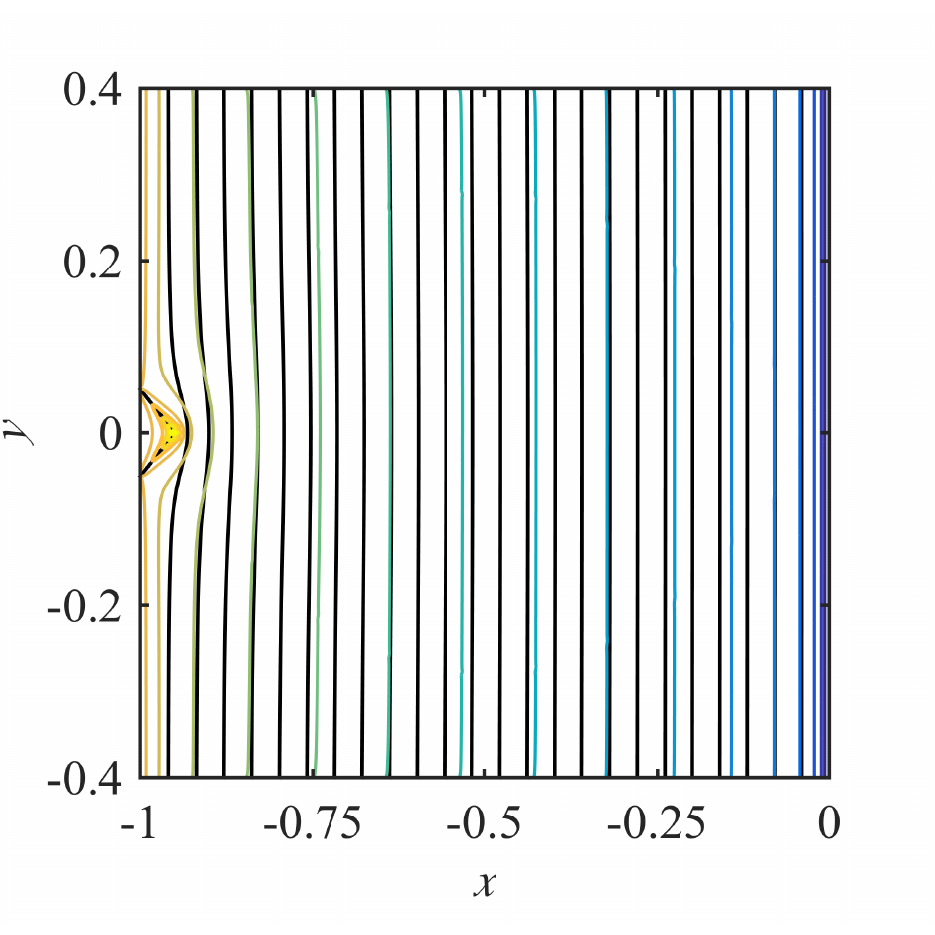}
  \put(-150,155){(a)}
  \label{fig:test1}
\end{minipage}%
\begin{minipage}{.5\textwidth}
  \centering
  \includegraphics[width=.74\linewidth]{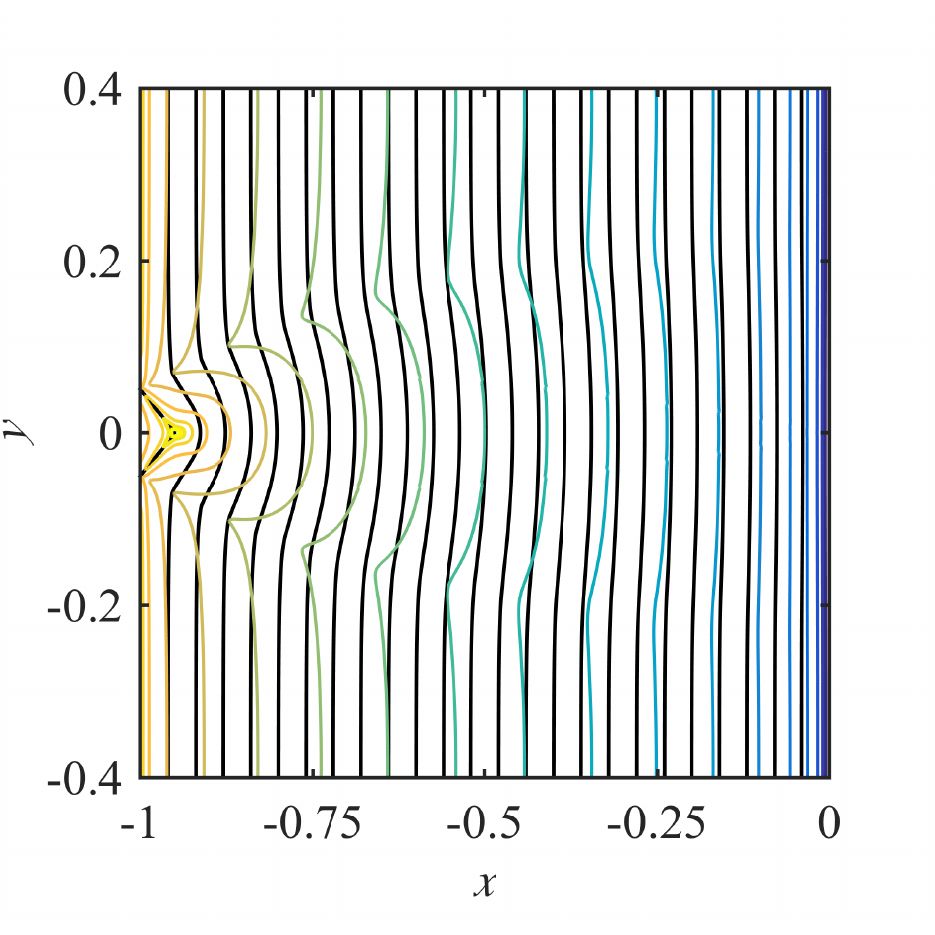}
  \put(-150,155){(b)}
  \label{fig:test1}
\end{minipage}%
\hfill
\begin{minipage}{.5\textwidth}
  \centering
  \includegraphics[width=.74\linewidth]{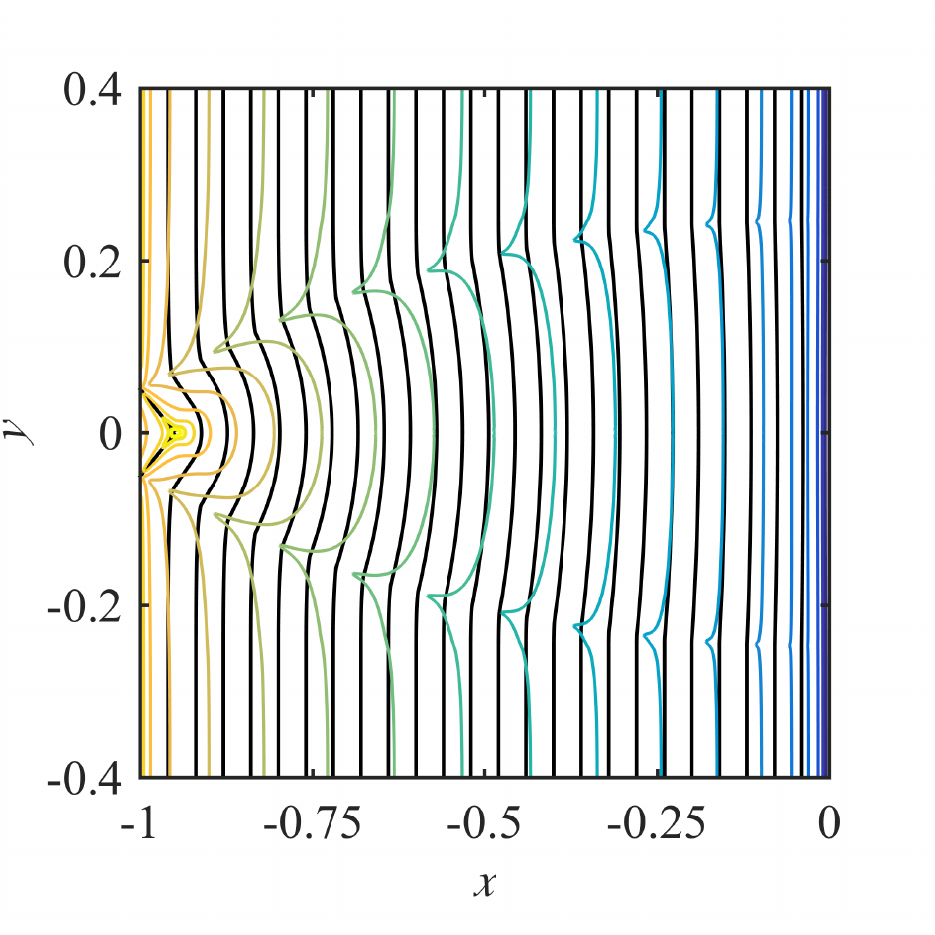}
  \put(-150,155){(c)}
  \label{fig:test1}
\end{minipage}%
\begin{minipage}{.5\textwidth}
  \centering
  \raisebox{0pt}{\includegraphics[width=0.74\linewidth]{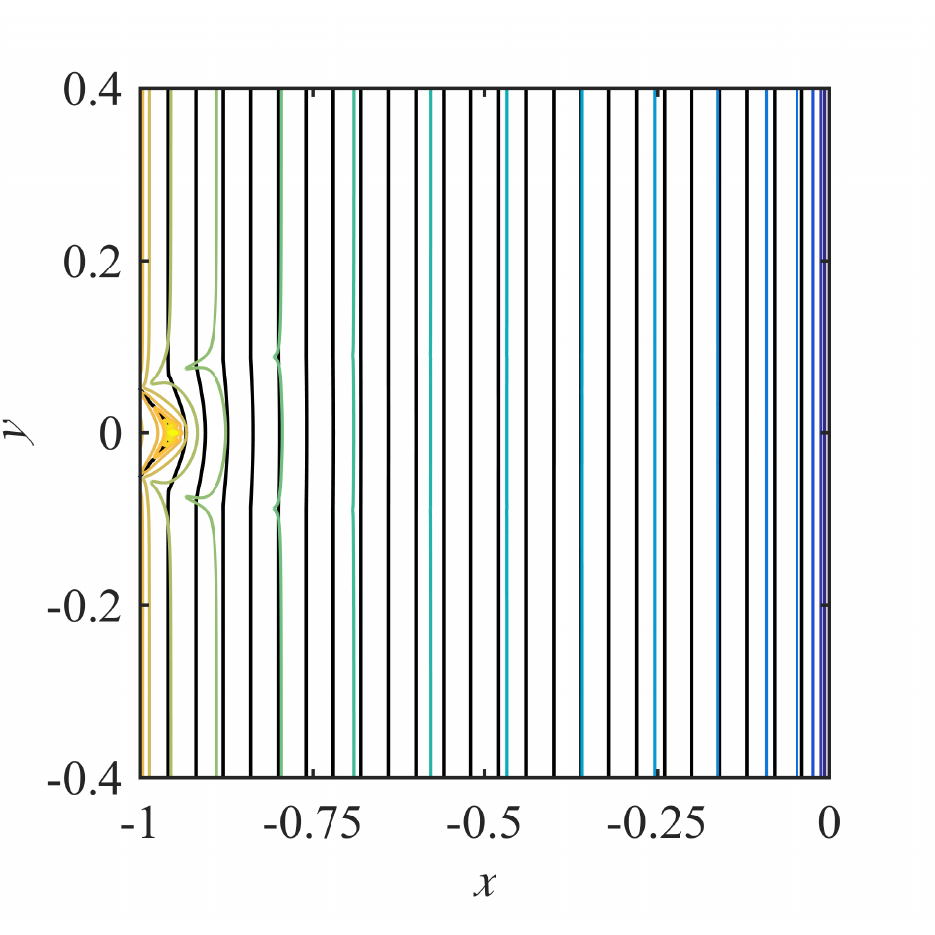}}
  \put(-150,155){(d)}
  \label{fig:test2}
\end{minipage}
\caption{Current streamlines in black and magnetic field contours in color in sample with triangular defect of height 0.05 and base 0.1 a) Ohmic case $n=1$, $r$ = 0.9957. b) $n=51$, $r$= 0.9593 c) Critical state or $n\rightarrow\infty$, $r$ = 0.95. d) Hybrid case with $n\rightarrow\infty$ for $J<J_0$ and $n=1$ for $J\geq J_0$, $r$ = 0.99999301. }
\label{FigGurf}
\end{figure*}

 The space coordinates $(x, y)$ associated to $\psi_i(J, \theta)$ can be obtained from the solution (\ref{z}) for the complex variable $z=x+iy$ in the different regions:

\begin{small}
\begin{multline}
\frac{z_1}{l}=-1+i\frac{a}{l}-e^{i\theta}\sum_{m=1}^{\infty} \frac{C_{4m}^{(1)}}{\tau_{4m}^+-1}\left( \frac{J}{J_0}\right)^{\tau_{4m}^+-1}\left[4m\cos (4m\theta) \right.\\ \left.-i {\tau_{4m}^+} \sin(4m\theta)\right],
\label{z1}
\end{multline}
\end{small}
\begin{small}
\begin{multline}
\frac{z_2}{l}=-i\frac{(l-a)}{l}+\frac{4}{\pi}\frac{J_0}{J}e^{i\theta}+\\ -e^{i\theta}\sum_{m=1}^{\infty} \frac{D_{4m}^{(2)}}{\tau_{4m}^- -1}\left( \frac{J}{J_0}\right)^{\tau_{4m}^- -1}\left[4m\cos (4m\theta) \right.\\ \left.-i {\tau_{4m}^-} \sin(4m\theta)\right]\\-r\ e^{i\theta}\sum_{m=1}^{\infty} \frac{C_{4m}^{(2)}}{\tau_{4m}^+ -1}\left( \frac{J}{J_m}\right)^{\tau_{4m}^+ -1}\left[4m\cos (4m\theta)\right.\\ \left. -i {\tau_{4m}^+} \sin(4m\theta)\right],
\label{z2}
\end{multline}
\end{small}
\begin{small}
\begin{multline}
\frac{z_3}{l}=-\frac{(l-a)}{l}+\\-r\ e^{i\theta}\sum_{m=1}^{\infty} \frac{D_{4m-2}^{(3)}}{\tau_{4m-2}^- -1}\left( \frac{J}{J_m}\right)^{\tau_{4m-2}^- -1} \left[ \right.\\ \left.(4m-2)\cos ((4m-2)\theta+\frac{3\pi}{2})\right.\\ \left. -i {\tau_{4m-2}^-} \sin((4m-2)\theta+\frac{3\pi}{2}) \right.\left.\right],
\label{z3}
\end{multline}
\end{small}
Where the integration constants have been chosen to guarantee space continuity of $\psi$. 
The parameters are all dependent on $D_{4k-2}^{(3)}$, and those depend at the same time on $r$, that is found numerically by enforcing space continuity of the solutions.
In the limit $n\rightarrow\infty$ the coefficients $\tau_{m}^+ \rightarrow m^2$ and $\tau_{m}^- \rightarrow -\infty$. This greatly simplify the stream function calculation. In particular the hodograph region (3) disappears and $r=(l-a)/l$ due to current conservation. Then $C_{4m}^{(2)}={2}/{m\pi}$ from (\ref{contJm}) and $C_{4m}^{(1)}={2}(r^{(4m)^{2}}-1)/{m\pi}$ from (\ref{contJ0}). \\
The current streamlines, or the same the contour or equipotential lines of the stream function, can be seen for different $n$ values in FIG. \ref{FigGurf}. Those noticably change direction around the triangle constriction much far from the constriction than in the ohmic case. The non-local magnetic field $H_d$ given by Biot-Savart's law has a well behaved local minimum around the same area of current direction change, and as in \cite{Jerem1}, this minimum was used here as the \textit{d}-lines, that are  displayed in FIG. \ref{FigPar}. This approach is practical as in magneto-optical experiments it is the magnetic field that is measured rather than the current distribution. The field $H_d$ was calculated by a Fourier transform method \cite{Vest1}, namely, $H_d=\mathcal{F}^{-1}\left[k \mathcal{F}(\psi)/2\right] $, where $\mathcal{F}$ is the 2D Fourier transform, and $\bm k$ the wave-vectors associated to the domain discretization.  As it can be seen in FIG. \ref{FigPar}, the \textit{d}-lines are always narrower than the one corresponding to the Bean case. This is expected as in a stripe it is impossible to keep constant current density with constrictions present. Furthermore,  if $J> J_c$, as expected for hodograph region (3), a flux flow regime must be considered near the indentation, where for finite $n$ the current density $J$ diverges. 

\subsection{Flux Flow}
The previous hodograph calculation assumes that the power law resistivity is valid even for currents above the current $J_0$, as in refs. \cite{Gur1,Gur2,Gur3}. If we consider the flux flow resistance \cite{Kim1}, then a better representation of the resistivity is given by \cite{Vest1}:
\begin{equation}
\rho=\left\{\begin{array}{ll}
\rho_0 (J/J_{c})^{n-1}, & J\leq J_{c}\\
\rho_0 , & J> J_{c},\\
\end{array}
\right.
\label{eqresist}
\end{equation} 
Assuming that $J_0=J_c$ for this case, here it is combined the creep regime for regions with $J<J_0$, and ohmic regions for $J>J_0$ using $n=1$. Lets obtain the relationship for the series expansions coefficients for region (1) with arbitrary creep exponent $\tilde{n}$ with corresponding $\widetilde{\tau}_{m}^+=\frac{1}{2}(1-\tilde{n}+ \sqrt{(\tilde{n}-1)^2+4\tilde{n}m^2})$, along with region (2) and region (3) both with creep exponent $n$ and $\tau_{m}^\pm=\frac{1}{2}(1-n\pm \sqrt{(n-1)^2+4nm^2})$.
The continuity relationships  (\ref{contJ0}), (\ref{contJm}) and (\ref{contdJm}) stay unmodified. Then by enforcing $z_1(J_0)=z_2(J_0)$ it gets:

\begin{multline}
C_{4m}^{(1)}\frac{\widetilde{\tau}_{4m}^+}{\widetilde{\tau}_{4m}^{+}-1}=D_{4m}^{(2)}\frac{{\tau}_{4m}^-}{{\tau}_{4m}^{-}-1}+\\+C_{4m}^{(2)}r^{{\tau}_{4m}^+}\frac{{\tau}_{4m}^+}{{\tau}_{4m}^{+}-1}-\frac{8}{\pi}\frac{4m}{(4m)^2-1}
\label{contzJ0_ff}
\end{multline}
That reduces to (\ref{contdJ0}) when $\tilde{n}=n$. And by combining eqns. (\ref{contJ0}), (\ref{contJm}), (\ref{contdJm}) and (\ref{contzJ0_ff}) an equation can be obtained to calculate  $D_{4k-2}^{(3)}$. Here is taken the particular case with $\tilde{n}\rightarrow\infty$ and $n=1$. With that choice, the relationships for the coefficients are finally:

\begin{equation}
C_{4m}^{(1)}=-\frac{2}{m\pi}+D_{4m}^{(2)}+r^{{4m}}C_{4m}^{(2)}
\label{contJ0_ff}
\end{equation}

\begin{equation}
D_{4m}^{(2)}=r^{4m}C_{4m}^{(2)}
\label{contzJ0_ffp}
\end{equation}

\begin{equation}
C_{4m}^{(2)}=\frac{2}{m\pi}-D_{4m}^{(2)}r^{4m}+\frac{8}{\pi}\sum_{k=1}^{\infty}\frac{4mD_{4k-2}^{(3)}}{(4m)^2-(4k-2)^2}
\label{contJm_ff}
\end{equation}

\begin{equation}
C_{4m}^{(2)}=D_{4m}^{(2)}r^{4m}-\frac{8}{\pi}\sum_{k=1}^{\infty}\frac{D_{4k-2}^{(3)} (4k-2) }{(4m)^2-(4k-2)^2}
\label{contdJm_ff}
\end{equation}

\begin{figure}
\centering \includegraphics[scale=0.75]{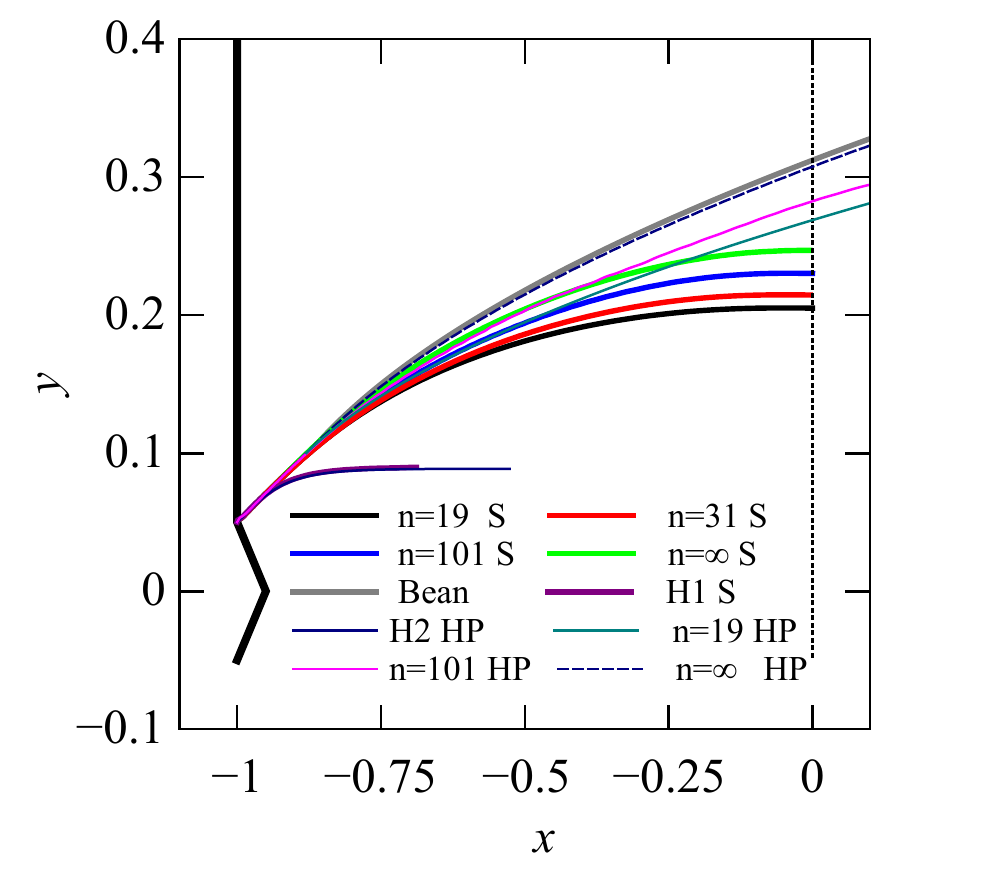}
\caption{Minimum in nonlocal magnetic field for different creep exponents in stripe (S) and infinite half plane (HP). Hybrid case (H1 and H2) correspond to  $n\rightarrow\infty$ for $J<J_0$ and $n=1$ for $J\geq J_0$. Values of $r$ parameter used are 0.9716, 0.9648, 0.95475, 0.95 and 0.99999301 for $n$=19 S, $n$=31 S, $n$=101 S, $n=\infty$ S  and H1 S respectively. Bean corresponds to Bean model parabola.}
\label{FigPar}
\end{figure}

Combing the above equations  (\ref{contzJ0_ffp}), (\ref{contJm_ff}) and (\ref{contdJm_ff}) it is obtained:

\begin{small}
\begin{equation}
\frac{r^{8m}-1}{4m}=\sum_{k=1}^{\infty}\frac{D_{4k-2}^{(3)}\left(4m(1-r^{8m})+(4k-2)(1+r^{8m})\right)}{(4m)^2-(4k-2)^2},
\label{eqD3s}
\end{equation}
\end{small}
that allows calculating $D_{4m-2}^{(3)}$ and then $D_{4m}^{(2)}$, $C_{4m}^{(2)}$ and $C_{4m}^{(1)}$. The $r$ parameter is again found numerically by enforcing continuity of the different solutions, and the current streamlines and magnetic field can be seen in FIG.\ref{FigGurf}(d). The \textit{d}-line for this case can also be found in FIG.\ref{FigPar}.
\section{Defect in infinite half plane}
A stream function for the infinite half plane $x\geq 0$ with a triangular defect that satisfies the boundary conditions is given by:
\begin{small}
\begin{equation}
\frac{\psi_1}{J_0}=\sum_{m=1}^{\infty} C_{4m}^{(1)}\left( \frac{J}{J_0}\right)^{\tau_{4m}^+}\sin (4m\theta),
\label{psi1hp}
\end{equation}
\end{small}
For $J<J_0$. And for $ J\geq J_0$:
\begin{footnotesize}
\begin{multline}
\frac{\psi_2}{J_0}=\sum_{m=1}^{\infty} D_{4m-2}^{(2)}\left( \frac{J}{J_0}\right)^{\tau_{4m-2}^-}\sin ((4m-2)\theta+\frac{3\pi}{2}),
\label{psi2hp}
\end{multline}
\end{footnotesize}

Notice the different units of distance of the $C$ and $D$ coefficients now. The $z$ functions are given by:
\begin{footnotesize}
\begin{multline}
{z_1}=-1+i{a}-{e^{i\theta}}\sum_{m=1}^{\infty} \frac{C_{4m}^{(1)}}{\tau_{4m}^+-1}\left( \frac{J}{J_0}\right)^{\tau_{4m}^+-1}\left[4m\cos (4m\theta) \right.\\ \left.-i {\tau_{4m}^+} \sin(4m\theta)\right],
\label{z1hp}
\end{multline}
\end{footnotesize}
\begin{footnotesize}
\begin{multline}
{z_2}=-(1-a)+\\ -{e^{i\theta}}\sum_{m=1}^{\infty} \frac{D_{4m-2}^{(2)}}{\tau_{4m-2}^- -1}\left( \frac{J}{J_0}\right)^{\tau_{4m-2}^- -1}\left[(4m-2)\cos ((4m-2)\theta+\frac{3\pi}{2}) \right.\\ \left.-i {\tau_{4m-2}^-} \sin((4m-2)\theta+\frac{3\pi}{2})\right],
\label{z2hp}
\end{multline}
\end{footnotesize}
Applying continuity conditions for $\psi$ and the terms involving the sine functions of $z$ it gets:
\begin{small}
\begin{equation}
D_{4m-2}^{(2)}=\frac{8}{\pi}\sum_{k=1}^{\infty}\frac{4kC_{4k}^{(1)}}{(4k)^2-(4m-2)^2}
\label{contJ0hp}
\end{equation}
\end{small}
\begin{multline}
\frac{{\tau_{4m-2}^-}}{\tau_{4m-2}^--1} D_{4m-2}^{(2)}=\frac{8}{\pi}\frac{a}{(4m-1)(4m-3)}+\\+\frac{8}{\pi}\sum_{k=1}^{\infty}\frac{4kC_{4k}^{(1)}}{(4k)^2-(4m-2)^2}\frac{\tau_{4k}^+}{{\tau_{4k}^+}-1}
\label{contzJ0hp}
\end{multline}
Combing eqns. (\ref{contJ0hp}) and (\ref{contzJ0hp}):
\begin{multline}
\sum_{k=1}^{\infty}\frac{4kC_{4k}^{(1)}}{(4k)^2-(4m-2)^2}\left(\frac{\tau_{4m-2}^-}{{\tau_{4m-2}^-}-1}-\frac{\tau_{4k}^+}{{\tau_{4k}^+}-1}\right)=\\\frac{a}{(4m-1)(4m-3)}
\label{ecChp}
\end{multline}
When $n\rightarrow\infty$ this expression simplifies to:
\begin{multline}
\sum_{k=1}^{\infty}\frac{4kC_{4k}^{(1)}}{(4k)^2-(4m-2)^2}\left(\frac{-1}{(4k)^2-1}\right)=\\\frac{a}{(4m-1)(4m-3)}
\label{ecChp}
\end{multline}
That can be better inverted as:
\begin{small}
\begin{multline}
C_{4k}^{(1)} =-((4k)^2-1)\times\\ \frac{64a}{\pi^2}\sum_{m=1}^{\infty}\frac{4k}{(4k)^2-(4m-2)^2}\frac{1}{(4m-1)(4m-3)}\\
=-\frac{32a}{\pi}k,
\label{ecCexplihp}
\end{multline}
\end{small}
by using the fact that 
\begin{footnotesize}
\begin{equation}
a_k=\left(\frac{8}{\pi}\right)^2 \sum_{m,l=1}^{\infty}\frac{4k}{(4k)^2-(4m-2)^2}\frac{4l}{(4l)^2-(4m-2)^2}a_l,
\end{equation}
\end{footnotesize}
obtained by applying eq. (\ref{ecMixSines}) to a expression of the kind {\small $\sum_{k=1}^{\infty} \sin(4k\theta)a_k=\sum_{m=1}^{\infty} \sin((4m-2)\theta+3\pi/2)b_m$}, and the sum in (\ref{ecCexplihp}) is calculated explicitly. The current streamlines and magnetic field can be seen in FIG.\ref{FigInfPlane}(a).
\subsection{Flux flow}
Again the case with creep exponent $\tilde n$ for $J<J_0$ and $n$ for $J\geq J_0$ is investigated, when $J_0=J_c$. Continuity conditions now produce:
\begin{small}
\begin{equation}
D_{4m-2}^{(2)}=\frac{8}{\pi}\sum_{k=1}^{\infty}\frac{4kC_{4k}^{(1)}}{(4k)^2-(4m-2)^2}
\label{contJ0hybhp}
\end{equation}
\end{small}
\begin{multline}
\frac{\tau_{4m-2}^-}{\tau_{4m-2}^--1} D_{4m-2}^{(2)}=\frac{8}{\pi}\frac{a}{(4m-1)(4m-3)}+\\+\frac{8}{\pi}\sum_{k=1}^{\infty}\frac{4kC_{4k}^{(1)}}{(4k)^2-(4m-2)^2}\frac{\widetilde\tau_{4k}^+}{{\widetilde\tau_{4k}^+}-1}
\label{contzJ0hybhp}
\end{multline}
And the equation for $C_{4k}^{(1)}$:
\begin{multline}
\sum_{k=1}^{\infty}\frac{4kC_{4k}^{(1)}}{(4k)^2-(4m-2)^2}\left(\frac{\tau_{4m-2}^-}{{\tau_{4m-2}^-}-1}-\frac{\widetilde\tau_{4k}^+}{{\widetilde\tau_{4k}^+}-1}\right)=\\\frac{a}{(4m-1)(4m-3)}
\label{ecCHybhp}
\end{multline}
And again for the hybrid case, a very high $\tilde n$ is considered for which the limit $\tilde n\rightarrow\infty$ is taken in the above expression, and $n=1$, which gives:
\begin{multline}
\sum_{k=1}^{\infty}\frac{4kC_{4k}^{(1)}}{(4k)^2-(4m-2)^2}\left(\frac{4m-2}{{4m-1}}-\frac{(4k)^2}{(4k)^2-1}\right)=\\\frac{a}{(4m-1)(4m-3)}
\label{ecCHybInfhp}
\end{multline}
The current streamlines and magnetic fields and \textit{d}-line can be found in FIG.\ref{FigInfPlane}(b) and FIG.\ref{FigPar} respectively.
\section{Discussion}
\begin{figure*}
\centering
\begin{minipage}{.5\textwidth}
  \centering
  \includegraphics[width=.74\linewidth]{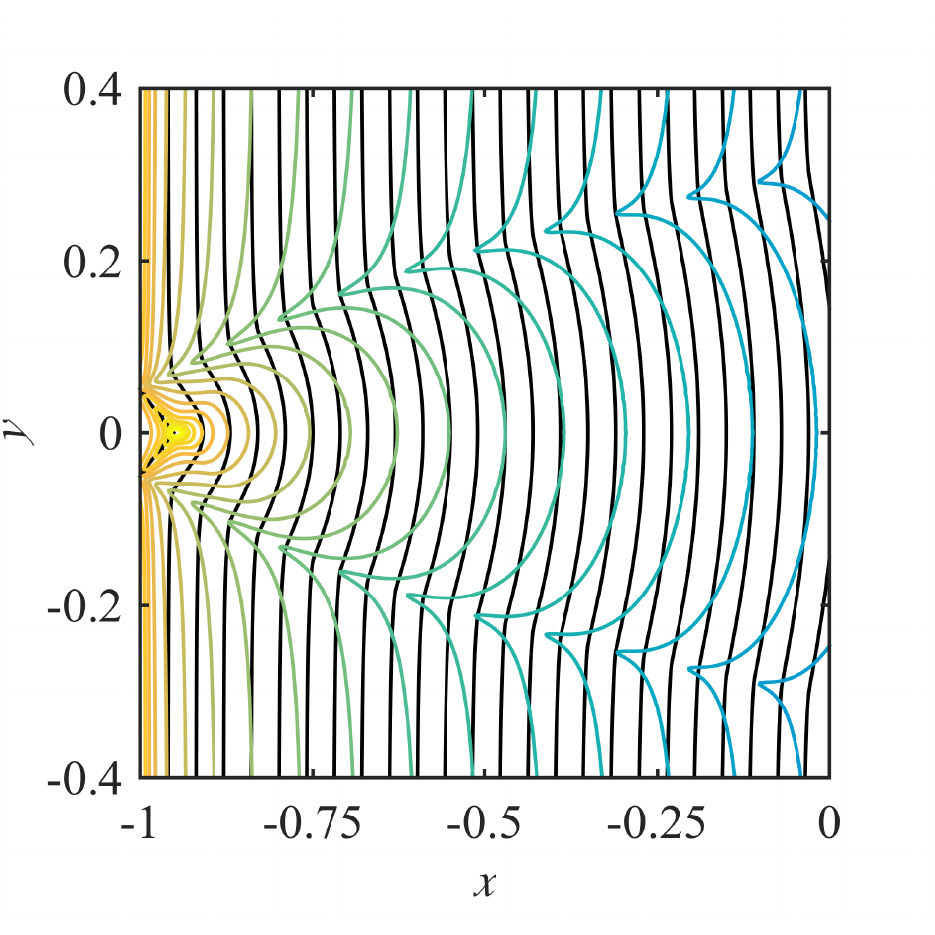}
  \put(-150,155){(a)}
  \label{fig:test1}
\end{minipage}%
\begin{minipage}{.5\textwidth}
  \centering
  \includegraphics[width=.74\linewidth]{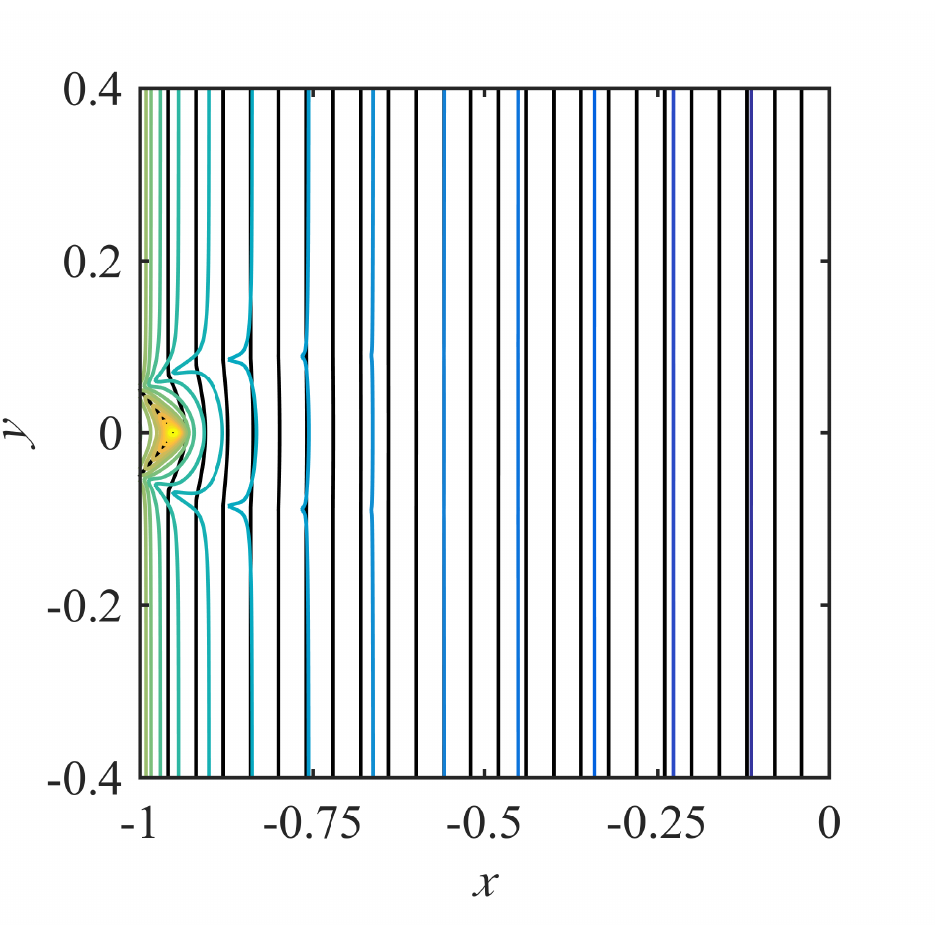}
  \put(-150,155){(b)}
  \label{fig:test1}
\end{minipage}%

\caption{Current streamlines in black and magnetic field contours in color in half infinite plane with triangular defect of height 0.05 and base 0.1 a) Critical state or $n\rightarrow\infty$. b) Hybrid case with $n\rightarrow\infty$ for $J<J_0$ and $n=1$ for $J\geq J_0$. Magnetic fields are calculated with currents in area twice the size shown.}
\label{FigInfPlane}
\end{figure*}

The triangular defect here studied shares several properties with the planar defect analyzed in \cite{Gur2}. In particular, the current density diverges on the tip of the triangle for finite $n$ due to the existence of the hodograph region (3), opposed to the critical state $n\rightarrow\infty$ in which the current density remains finite in that spot. For the infinite half plane it can be verified that $J=J_0$ at the triangle tip, by evaluating $z_1(\theta=3\pi/4-\delta, J=J_0)$ from equation (\ref{z1hp}) with the aid of expression (\ref{ecCexplihp}) for $C_{4k}^{(1)}$:
\begin{small}
\begin{multline}
{z_1}=-1+i{a}-{e^{i(\frac{3\pi}{4}-\delta)}}\frac{8a}{\pi}\sum_{m=1}^{\infty} \frac{{4m}^2}{{4m}^2-1}(-1)^m \cos(4m\delta)+\\
 +i{e^{i(\frac{3\pi}{4}-\delta)}}\frac{8a}{\pi}\sum_{m=1}^{\infty} \frac{{4m}^3}{{4m}^2-1}(-1)^m \sin(4m\delta),
\end{multline}
\end{small}
And by taking the limit $\delta\rightarrow0$ it is obtained $z_1=-1+a$, probing that $J=J_0$ there. In the case of the critical state in the stripe, it can be equally calculated that $J=J_m$ at the tip of the triangle, the same as on the opposite side of the stripe. In fact $J=J_m$ on all the line connecting both points, since $J=J_m$ on both sides of the constriction, and all the current is passing through a zone where the maximum $J$ is $J_m$. It also follows that $r=(l-a)/l$.\\ 
 In the hybrid case for the stripe, most of the current significantly above $J_0$ passes near the triangle tip, and the current density $J$ is very close to $J_m$ in most of the constriction. Additionally, $J_m$ is also very close to $J_0$,  as it can be seen from the $r=0.99999301$ value. 
 \\In the case of the infinite half plane with hybrid creep exponents, it is seen that perturbations on the currents are only important near the indentation, similar to the purely ohmic case, virtually disappearing at distances larger than $\sim 10a$, as it can be seen in the current streamlines in FIG.\ref{FigGurf}(d)  and their \textit{d}-lines in FIG.\ref{FigInfPlane}(b). The \textit{d}-lines for the stripe and infinite plane also almost identical in that case. 
 \\The \textit{d}-line in the critical state for the infinite half plane traces a curve quite close to the Bean parabola, but identical actually to the result obtained in \cite{Gur2} of $x\approx y^2/1.94a$ for the planar defect. In the appendix it is shown that both the planar defect and triangular indentation have actually the same asymptotic \textit{d}-line. This is done by replacing the expression (\ref{ecCexplihp}) for  $C_{4k}^{(1)}$, obtained here just from continuity conditions, opposed to the calculation for the planar defect \cite{Gur2} where $C_{4k}^{(1)}$ is obtained from a expansion of current vortices and anti-vortices. The same expansion procedure is also done in the appendix to obtain approximate explicit expressions for the triangle indentation $C_{4k}^{(1)}$ and $D_{4m-2}^{(2)}$ coefficients, that have similar functional form\cite{Gur2} and limit for high $n$ as the planar defect. A key point of the derivation is that the behaviour of the currents is mainly determined by a vortex solution to a London equation in a transformed hodograph plane of coordinates $\eta=(1/\sqrt{n})\log(E/E_0)$ , $\theta$. That vortex is centred in $\eta=0$, $\theta=\pi/2$, and is the main contributor to the Fourier expansions of $\psi$ in both the planar defect and in the triangle indentation. Here both defects share only the height $a$, giving the same far current behavior, a feature already expected in the Bean model\cite{Jerem1}, where the asymptotic \textit{d}-line depends only on the height of the defect.

\section{Conclusion}
The hodograph series method is used to solve the problem of currents circulating in a straight stripe of finite width and an infinite half plane with a triangular border indentation, showing the non-local magnetic fields produced by those currents. Both cases present similar properties as the planar defect already discussed in \cite{Gur2, Gur3}, with diverging currents in the defect tip for finite $n$, and bounded current for the critical state. An hybrid case of flux-flow and creep was also considered, showing the presence of \textit{d}-lines, but of short range. For the triangular defect in the infinite half plane, the same asymptotic \textit{d}-line is obtained as that of the planar defect in the critical state, which is explained here by the similar contribution of vortex currents in a modified hodograph space $\eta$, $\theta$. 

\section{Acknowledgements} Grants \textsc{ARC}  13/18-08 for Concerted Research Actions, financed by the Wallonia-Brussels Federation, and Batwal D8.31404.060-P. J. I. A. also acknowledges \textsc{FNRS} support reference 2016/V 3/5/013 (W) -IB/LS-6967.

\appendix*
\section{Asymptotics of currents for infinite half plane}
As per \cite{Gur1,Gur2,Gur3}, eq. (\ref{psi}) is the solution of the equation appearing from the hodograph transformation:
\begin{equation}
\frac{J^2}{n}\frac{\partial ^2\psi}{\partial J^2}+J\frac{\partial \psi}{\partial J}+\frac{\partial ^2\psi}{\partial \theta^2}=0
\label{ecJHod}
\end{equation}
Whereas the transforming expressions for space coordinates (\ref{z}) are obtained from the equations:
\begin{align}
\frac{\partial z}{\partial E}=-e^{i\theta}\left(\frac{E\sigma}{J^2}\frac{\partial \psi}{\partial \theta}-i\rho\frac{\partial \psi}{\partial E}\right)\label{dzdE}\\
\frac{\partial z}{\partial \theta}=-e^{i\theta}\left(-\frac{E^2}{J}\frac{\partial \psi}{\partial E}-i\rho\frac{\partial \psi}{\partial \theta}\right)\label{dzdtheta}
\end{align}

Where $\sigma=\partial J/\partial E$. The following change of variable can be done for (\ref{ecJHod}):
\begin{align}
\eta&=(1/\sqrt{n})\log(E/E_0)=\sqrt{n}\log(\frac{J}{J_0})\\
\psi&=e^{-\beta\eta}h
\end{align}
With $\beta=\frac{n-1}{2\sqrt{n}}$, and eq.(\ref{ecJHod}) transforms into a London equation:
\begin{equation}
\frac{\partial ^2h}{\partial \eta^2}+\frac{\partial ^2 h}{\partial \theta^2}-\beta^2 h=0
\label{ecLondon}
\end{equation}
Considering $\eta$ and $\theta$ as cartesian coordinates, a change of variable to polar coordinates $r$, $\chi$ is performed, with $\eta=r\cos \chi$, $\theta=\theta_0+\sin \chi$. A vortex solution is obtained in these polar coordinates:
\begin{equation}
h=\sqrt{2}h_0\frac{e^{-\beta\eta}}{\sqrt{r}}\cos \frac{\chi}{2}
\label{ecVortex}
\end{equation}
The solution for $\psi$ can be constructed by adding different vortices of the same strength on the line $\eta=0$ and alternating signs such that $\psi=0$ at $\theta=\frac{\pi}{4}, \frac{3\pi}{4},...$. This is done by adding terms of the form (\ref{ecVortex}) with sign $(-1)^{m+1}$ centred at points $\theta_m=\frac{m\pi}{2}$, and converting back to the $\eta$, $\theta$ coordinates:
\begin{equation}
\psi=h_0\sum_m (-1)^{m+1}\frac{\sqrt{r_m+\eta}}{r_m}e^{-\beta(r_m+\eta)}
\label{ecPsiVortex}
\end{equation}
Where $r_m=\sqrt{\eta^2+(\theta-m\pi/2)}$. The coefficients  $C_{4k}^{(1)}$ can be obtained then by calculating the Fourier terms:
\begin{equation}
C_{4k}^{(1)}=\frac{8}{\pi}\int_{\pi/2}^{3\pi/4}\psi(J_0,\theta)\sin(4k\theta)d\theta
\end{equation}
The main contributor to that integral is the term with $m=1$ of (\ref{ecPsiVortex}), so the other $m$ terms are discarded. For $n \gg 1$, $\psi(J_0,\theta)$ is very peaked around $\theta=\pi/2$ so the upper limit can be extended to infinite, and the following expressions are obtained:
\begin{small}
\begin{equation}
C_{4k}^{(1)}=h_0\frac{8}{\sqrt{2\pi}}\frac{\sqrt{\sqrt{\beta^2+(4m)^2}-\beta}}{\sqrt{\beta^2+(4m)^2}}
\label{ecCvortex}
\end{equation}
\end{small}
\begin{small}
\begin{equation}
D_{4m-2}^{(2)}=h_0\frac{8}{\sqrt{2\pi}}\frac{\sqrt{\sqrt{\beta^2+(4m-2)^2}+\beta}}{\sqrt{\beta^2+(4m-2)^2}}
\label{ecDvortex}
\end{equation}
\end{small}
Where the same principle was used to obtain (\ref{ecDvortex}). 
The parameter $h_0$ is obtained by comparing the behaviour of (\ref{ecCvortex}) for high $n$:
\begin{equation}
C_{4m}^{(1)}\approx h_0 \frac{32m}{n^{3/4}}\sqrt{\frac{2}{\pi}},
\end{equation}
to the value already calculated here in (\ref{ecCexplihp}) from continuity conditions for $n\rightarrow\infty$ and equal to $-32ma/\pi$, then:
\begin{equation}
h_0=-\frac{n^{3/4}a}{\sqrt{2\pi}}
\end{equation}
The \textit{d}-line can be obtained from the asymptotic behaviour of the current, that can be calculated by adding up the series (\ref{z1hp}) when replacing (\ref{ecCexplihp}), for $n\rightarrow \infty$:
\begin{small}
\begin{multline}
{z_1}=-1+i{a}+\frac{8a}{\pi}{e^{i\theta}}\sum_{m=1}^{\infty} \frac{4m}{(4m)^2-1}\left( \frac{J}{J_0}\right)^{(4m)^2-1}\times\\ \left[4m\cos (4m\theta)  -i {(4m)^2} \sin(4m\theta)\right],
\end{multline}
\end{small}
This expression can be written as a function of the series $I_0$:
\begin{equation}
{z_1}=-1+i{a}+\frac{8a}{\pi}{e^{i\theta}}\left(I_0+i\frac{\partial I_0}{\partial \theta}\right)
\label{z1I0}
\end{equation}
With:
\begin{small}
\begin{align}
I_0=&\sum_{m=1}^{\infty} \frac{(4m)^2}{(4m)^2-1}\left( \frac{J}{J_0}\right)^{(4m)^2-1}\cos (4m\theta) \\
=&\sum_{m=1}^{\infty} \left(1+\frac{1}{(4m)^2-1}\right)\left( \frac{J}{J_0}\right)^{(4m)^2-1}\cos (4m\theta) 
\label{ecI0}
\end{align}
\end{small}
Then, for $\tau=\log(J_0/J)>0$, $z\gg a$, $J\approx J_0$, $\tau \ll 1$, it gets $\left({J}/{J_0} \right)^{(4m)^2-1}=e^{-\tau((4m)^2-1)}\approx e^{-\tau (4m)^2}$ and:
\begin{equation}
I_0\approx\sum_{m=1}^{\infty} \left(1+\frac{1}{(4m)^2-1}\right)e^{-\tau (4m)^2}\cos (4m\theta) 
\end{equation}
\begin{equation}
I_0\approx\sum_{m=1}^{\infty} e^{-\tau (4m)^2}\cos (4m\theta) +\sum_{m=1}^{\infty}\frac{e^{-\tau (4m)^2}}{(4m)^2-1}\cos (4m\theta) 
\end{equation}
The exponential term in the second sum can be omitted for $\tau \ll 1$, and replacing the angle $\gamma=\theta-\pi/2$:
\begin{equation}
I_0\approx\sum_{m=1}^{\infty} e^{-\tau (4m)^2}\cos (4m\gamma) +\sum_{m=1}^{\infty}\frac{\cos (4m\gamma) }{(4m)^2-1}
\end{equation}
The first sum can be estimated with the Euler-Maclaurin formula \cite{Abramowitz}, giving $\exp(-\gamma^2/4\tau)\sqrt{\pi}/8\sqrt{\tau}-1/2$ and the second adds to $1/2-\pi \sin(2\gamma)/4$. Then:
\begin{equation}
I_0\approx\frac{\sqrt{\pi}}{8\sqrt{\tau}}e^{-\gamma^2/4\tau}-\frac{\pi}{4}\sin 2\gamma
\end{equation}
\begin{equation}
I_0^\prime=\frac{\partial I_0}{\partial \theta}\approx-\frac{\sqrt{\pi}\gamma}{16{\tau}^{3/2}}e^{-\gamma^2/4\tau}-\frac{\pi}{2}\cos 2\gamma
\end{equation}
Then, for $y\gg a$, $x\gg 1$, from (\ref{z1I0}):
\begin{align}
x&=b\left(I_0 \sin\gamma +I_0^\prime \cos\gamma\right)\\
y&=b\left(-I_0 \cos\gamma +I_0^\prime \sin\gamma\right)
\end{align} 
With $b=8a/\pi$. By replacing $I_0$ and $I_0^\prime$, naming $e_1=\exp(-\gamma^2/4\tau)\sqrt{\pi}/8\sqrt{\tau}$:
\begin{align}
x&=be_1(\sin\gamma +\frac{\gamma}{2\tau}\cos\gamma)\label{xe1}\\
y&=be_1(\cos\gamma -\frac{\gamma}{2\tau}\sin\gamma)\label{ye1}
\end{align} 
By neglecting the terms not proportional to $e_1$. Then, eliminating $e_1$, and solving for $\tau$:
\begin{equation}
\tau=-\frac{\gamma}{2}\frac{(x\sin\gamma-y\cos\gamma)}{(x\cos\gamma+y\sin\gamma)}\label{tau(x,y)}
\end{equation}
Combining (\ref{xe1}) and (\ref{ye1}), replacing back in (\ref{tau(x,y)}), and expanding for $\gamma\ll 1$:
\begin{equation}
\frac{32}{\pi b^2}x^2=\frac{(1+\xi\gamma)}{(\xi-\gamma)^3{\gamma}}{e^{-\gamma\frac{(1+\xi\gamma)}{(\xi-\gamma)}}}
\label{Paramxygammaxi}
\end{equation}
In which $\xi=y/x$. As discussed in \cite{Gur2}, the \textit{d}-line can be obtained from the line where the space derivatives of $\gamma(x,y)$ diverge. This can be calculated from the condition $\partial x/\partial \gamma =0$. Then (\ref{Paramxygammaxi}) is approximated for $\xi\ll 1$,  derived with respect to $\gamma$ while keeping $\xi$ constant and equalled to zero, from which it is obtained $\gamma=\xi/2$. Replacing this back in (\ref{Paramxygammaxi}) and retaining the lowest order in $\xi$, the expression  $x=y^2/(8a/\sqrt{2\pi e})$ for the \textit{d}-line is obtained.


\section*{References} 

\begin{thebibliography}{99}


\bibitem{Brandt0} E. H. Brandt , Rep. Prog. Phys. {\bf 58}, 1465 (1995).

\bibitem{Koshelev} A. E. Koshelev, I. A. Sadovskyy, C. L. Phillips, A. Glatz, Phys. Rev. B \textbf{93}, 060508(R) (2016).

\bibitem{Blatter} G. Blatter, V. B. Geshkenbein, J. A. G. Koopmann, Phys. Rev. Lett. \textbf{92}, 067009 (2004).


\bibitem{Jerem1} J. Brisbois, O.-A. Adami, J. I. Avila, M. Motta, W. A. Ortiz, N. D. Nguyen, P. Vanderbemden, B. Vanderheyden, R. B. G. Kramer, A. V. Silhanek, Phys. Rev. B \textbf{93}, 054521 (2016).

\bibitem{Mints} R. G. Mints, E. H. Brandt, Phys. Rev. B \textbf{54}, 12421 (1996).

\bibitem{Bean1} C. P. Bean, Phys. Rev. Lett. \textbf{8}, 250 (1962)

\bibitem{Bean2} C. P. Bean, Rev. Mod. Phys. \textbf{36}, 31 (1964 ).

\bibitem{Bean3} C. P. Bean, J. Appl. Phys. \textbf{41}, 2482 (1970).

\bibitem{Schuster1} Th. Schuster, M. V. Indenbom, M. R. Koblischka, H. Kuhn, H. Kronmüller, Phys. Rev. B {\bf 49}, 3443 (1994).

\bibitem{Schuster2} Th. Schuster, H. Kuhn, M. V. Indenbom, Phys. Rev. B \textbf{52}, 15621 (1995).

\bibitem{Eisenmenger} J. Eisenmenger, P. Leiderer, M. Wallenhorst, H. Dötsch, Phys. Rev. B \textbf{64}, 104503 (2001).

\bibitem{Vinokur1} V. M. Vinokur , M. V. Feigel'man, V. B. Geshkenbein,  Phys. Rev. Lett. {\bf 67}, 915 (1991).

\bibitem{Rak} A. L. Rakhmanov, D. V. Shantsev, Y. M. Galperin, T. H. Johansen, Phys. Rev. B \textbf{70}, 224502 (2004).

\bibitem{Deni} D. V. Denisov, A. L. Rakhmanov, D. V. Shantsev, Y. M. Galperin, T. H. Johansen, Phys. Rev. B \textbf{73}, 014512 (2006).

\bibitem{Gur1} A. Gurevich, J. McDonald, Phys. Rev. Lett. \textbf{81}, 2546 (1998).

\bibitem{Gur2} A. Gurevich, M. Friesen, Phys. Rev. B \textbf{62}, 4004 (2000).

\bibitem{Gur3} M. Friesen, A. Gurevich, Phys. Rev. B \textbf{63}, 064521 (2001).

\bibitem{Camp1} A. M. Campbell, J. E. Evetts,  \textit{Critical Currents in Superconductors}, Monographs on Physics. Taylor \& Francis Ltd., London (1972).

\bibitem{Kim1} Y. B. Kim, C. F. Hempstead, A. R. Strnad, Phys. Rev. \textbf{139}, A1163 (1965).


\bibitem{Vest1} J. I. Vestg{\aa}rden , P. Mikheenko, Y. M. Galperin, T. H. Johansen, New J. Phys. {\bf15}, 093001 (2013).

\bibitem{Abramowitz} Formula 3.6.28 in \textit{Handbook of Mathematical Functions With Formulas, Graphs, and Mathematical Tables},  by M. Abramowitz and I. A. Stegun, National Bureau of Standards, Applied Mathematics Series - 55, 1972. 


\end{thebibliography}
\end{document}